\documentstyle[12pt]{article}
\begin{document}
\titlepage
 \begin{flushright}
USP-IFQSC/TH/93-05 \end{flushright} \vspace{0.5cm}
\title{ Exact solution of the deformed biquadratic spin 1 chain}
\author{    ROLAND K\"{O}BERLE
             \thanks{ Supported in part by CNPq-BRASIL.} \\
              Instituto de F\'{\i}sica e Qu\'{\i}mica
 de S\~ao Carlos,
              Universidade de S\~ao Paulo\\
               Caixa Postal 369, S\~ao Carlos 13560, Brasil \\
and \\
          A. LIMA - SANTOS *\\
             Departamento de F\'{i}sica, Universidade
Federal de S\~ao Carlos\\
             Caixa Postal 676, S\~ao Carlos, Brasil
       }
\maketitle
\begin{abstract}
 We solve for the spectrum of the deformed biquadratic spin 1
 chain by a generalization of the coordinate Bethe-Ansatz.
The model is invariant under the quantum group $U_q sl(2)$.
 We consider several boundary conditions and find in particular
that for free boundary conditions the
Bethe states are highest weight states of the
quantum group. Depending on the deformation parameter $q$,
the models's hamiltonian may not be hermitian,  but it
always contains the complete spectrum of the $6$-vertex
 model.
\end{abstract}
\newpage
%
%% FOLLOWING LINE CANNOT BE BROKEN BEFORE 80 CHAR
%%******************************************************************************

A model possessing an infinite number of conservation laws is
 usually referred to as {\em exactly soluble}, even if there
is no way in sight to in fact obtain at least a partial
solution. Such exactly soluble models can be generated in
two dimensions by solving Yang-Baxter, star-triangle or
similar cubic equations\cite{YB}, which guarantee the
existence of an infinite number of commuting conserved
 charges.

One of the favorite methods to apply in such a case is
 the {\em Bethe-Ansatz} \mbox{( BA )}.
The more sophisticated form is the {\em algebraic} BA\cite{TF}.
 Here one uses the miraculous fact, that the Yang-Baxter
equations can be recast in the form of commutation relations
for creation and destruction operators with respect to a
convenient reference state. It is not known, when a particular
 solution of the Yang-Baxter equations can be used as starting
point for the algebraic BA.

The simplest version is the {\em coordinate } BA\cite{B} and
we will use  a generalization of this framework  to
{\em  solve} the deformed biquadratic spin $1$ hamiltonian
subjected to several types of boundary conditions.
This hamiltonian has been obtained as a solution of the
Yang-Baxter equations for spin $1$ with
$U(1)$-invariance\cite[AKL]. Yet the Yang-Baxter equations
do not in this case provide one with commutations
relations, so that the algebraic version of the
BA cannot readily applied.

This model has been studied in its undeformed version by
Parkinson\cite{P}, who obtained the ground state and
several low energy excitations.
Barber \& Batchelor\cite{BB} have shown that the
hamiltonian satisfies the Temperley-Lieb
algebra\cite{TL}.
Kl\"umper has obtained the energy gap using
inversion relations\cite{K}.

The deformed model was introduced by Batchelor et
al.\cite{BMNR}, who considered spin $1$ hamiltonians
invariant under the quantum group $U_q sl(2)$. It is one
of the cases, which cannot be obtained by a fusion
procedure from integrable spin $1/2$
hamiltonians\cite{KR}, so that this method of solution is not
available.

In terms of spin one variables $\vec{S}_k=(S_k^x,S_k^y,S_k^z)$
 the deformed biquadratic spin $1$ hamiltonian is given by:
\begin{equation}
     H(q)=\sum_{k=1}^N e_{k},
\end{equation}
with
\begin{displaymath}
  e_{k} = \left( \vec{S}_k\cdot \vec{S}_k\right)^2 - 1 -
             \sinh^2\lambda\left[ S^z_k S^z_{k+1} -
              (S^z_k S^z_{k+1})^2)\right]
\end{displaymath}
\begin{displaymath}
             +\frac{1}{2}\sinh\lambda\left[
             (S_k^x S_{k+1}^x + S_k^y S_{k+1}^y)
             (S_{k+1}^z - S_k^z ) + h.c.\right]
\end{displaymath}
\begin{equation}
          + 2\sinh^2(\lambda/2)\left[
             (S_k^x S_{k+1}^x +
        S_k^y S_{k+1}^y) S_k^z S_{k+1}^z + h.c.
                 \right]
\end{equation}
\begin{displaymath}
          +\frac{1}{2}\sinh(2\lambda)
      \left[ S_k^z S_{k+1}^z (S_{k+1}^z - S_k^z)
                                \right],
\end{displaymath}
where $ \vec{S}_k\cdot \vec{S}_{k+1}$ denotes the standard
rotationally invariant scalar product and $\lambda$ is
 a coupling constant. In the sequel we will also use
the deformation parameters $q$ and $\beta$:
$\beta=e^{2\lambda},
q+1/q=1+\beta+1/\beta$. At $\lambda=0$ $e_k$ reduces
 to the isotropic spin $1$ biquadratic chain\cite{BB,P}.
%\begin{equation}
%   e^{(3)}(\beta) =[11|33]\,\hat{q}^2 + [33|11]/\hat{q}^2 +
%         [22|22] +  [13|31] + [31|13] +%
%\label{TL}
%\end{equation}
%\begin{displaymath}
%                    \left( [12|32]+[21|23]\right) \hat{q} +
%                    \left( [23|21]+[32|12]\right) /\hat{q},
%\end{displaymath}
%where now $\beta =\hat{q}^2+1/\hat{q}^2+1$.

$e_k$ can be obtained\cite{AKL} as a solution of the spin
one representation of the Hecke algebra with only the
$U(1)$-symmetry ( generated by
\mbox{ $\sum_i S_z(i)$.}). In fact $e_k$ obeys the more
 restrictive Temperley-Lieb algebra:
\begin{displaymath}
        e_k^2  =  (q+1/q)e_k,\;\;e_k e_{k+1} e_k=e_k,\;\;
\end{displaymath}
\begin{equation}
    [e_k,e_l]=0,\;\;|k-l|\geq 2.
\end{equation}
 and  commutes with the quantum group
$U_q sl(2) $: the hamiltonian  density is the Casimir
operator
 $\left ( {\cal S}_k\cdot{\cal S}_{k+1}\right )^2$,
where ${\cal S}$ is the spin 1 representation of the
$U_q sl(2) $ algebra for one site.

% ******************  The coordinate BA. ********************

In order to diagonalize $H$, we proceed as in usual BA
applications. Due to $U(1)$-invariance, there always exists
 a reference state $|\Omega>$, such that $H|\Omega> = 0$.
In the basis, where $S_k^z$ is diagonal with eigenvectors
 $|+,k>,|0,k>$ and $|-,k>$ and eigenvalues $+1,0$ and $-1$
respectively, we take $|\Omega>$ to be
$|\Omega>=\prod_k |+,k>$. This is the only eigenstate in
 the sector $r=0$, where we label the sectors by the eigenvalues
 of $r = (N - \sum_i S_k^z)/2$. For simplicity we will in this
paper only discuss the more interesting case, where the states
$|0,i>$ occur in neighboring pairs, so that they move\footnote{
   Single $|0>$ sorrounded by $|+>$ do not move under the action
of $H$ and  behave rather like stationary impurities.}.
Therefore $r$ will be an integer.

In the sector $r=1$, eigenstates $|r=1>$ will be a
superposition of the states
$|x_{\,-}>=|+,1>\ldots|-,x>\ldots|+,N>$ and
$|x_{\,0}>=|+,1>\ldots|0,x>|0,x+1>\l
   dots|+,N>$, or
 in pictures:
\begin{eqnarray}
    |x_{\,-}>=  & (+++\ldots &- \ldots +++) \nonumber\\
            &            &x              \nonumber\\
    |x_{\,0}>=  & (+++\ldots &00 \ldots +++)  \nonumber\\
            &            &x                    \nonumber
\end{eqnarray}
i.e. $|r=1>=\sum_x a(x)|x_{\,-}> + \;b(x) |x_{\,0}>$.
 To start let us choose periodic boundary conditions
 maintaining translational invariance:
$\vec{S}(N+1)=\vec{S}(1)$, implying $a(x+N)=a(x),b(x+N)
=b(x)$.
Thus we parametrize as: $a(x)=a\exp(\imath\theta x)$
and $b(x)=b\exp(\imath\theta x)$,
 $\theta$ being the momentum.
When $H$ now acts on $|r=1>$ it sees the reference
configuration, except in the vicinity of $x$ and we
obtain the eigenvalue equations\footnote{
        In normal applications of the BA, $H$ is diagonal
 in this basis.}:
\begin{eqnarray}
       E a(x) = & (\beta +1/\beta) a(x) +a(x+1) +a(x-1) +
\sqrt\beta b(x-1)+1/\sqrt\beta b(x)
              \nonumber\\
       E b(x) = & b(x) +\sqrt\beta a(x+1) +
1/\sqrt\beta a(x).
\end{eqnarray}
This system yields two eigenstates $|1>,|2>$ and
their energies:
\begin{equation}
\begin{array}{llll}
   |1> & =\sum_x e^{\imath\theta x} [\gamma(-\theta)|x_{\,-}>
 + \;|x_{\,0}>], \;\;
  & E_1 & =1+\gamma(\theta)\gamma(-\theta)\\
   |2> & =\sum_x e^{\imath\theta x} [|x_{\,-}> -
\;\gamma( \theta)|x_{\,0}>], \;\;
  & E_2 & = 0,
\end{array}
\end{equation}
where $\gamma(\theta)=1/\sqrt\beta+e^{\imath\theta}
\sqrt\beta$ and $\theta$ being determined by the
periodic boundary condition:
\begin{equation}
            e^{\imath N\theta} = 1.
\end{equation}
%Notice the symmetry $E(\theta)=E(-\theta)$, which follows
 from parity %invariance and will be crucial for our procedure
 to work for free bc.
We describe this situation by saying that we have two
types of pseudoparticles with energies $E_1$ and $E_2$.

The situation becomes  nontrivial for $r>1$, where we can
 have interacting pseudoparticles.
The main result of this paper is to show, that $H$ can be
 diagonalized in a convenient basis, constructed from
products of single pseudoparticle eigenfunctions. From
this statement, it  immediately follows that the eigenvalues
 of $H$ will be a sum of single pseudoparticle energies:
\begin{equation}
        E = \sum_{n=1}^r E_n  =
   \sum_{n=1}^r \epsilon_n \left [ 1+\gamma(\theta_n)
\gamma(-\theta_n)
                                  \right ],
\end{equation}
where $\epsilon_n$ depends on which type of pseudoparticle
 we use:
$\epsilon_n=1$ for $E_n\neq 0$ and $\epsilon_n=0$ for $E_n=0$.
$\theta_n$ are rapidities still undetermined at this stage.

Since all the interesting information can be extracted from
the sector $r=2$, we will treat this case in some detail.
 We seek its eigenstates in the form :
\begin{equation}
|\theta_1,\theta_2> = \sum_{x<y} a(x,y)|x_{\,-}>
|y_{\,-}> +\;  b_1(x,y)|x_{\,-}>|y_{\,0}>
\label{2par}
\end{equation}
\begin{displaymath}
     +\; b_2(x,y)|x_{\,0}>|y_{\,-}> +\;
           c(x,y)|x_{\,0}>|y_{\,0}>
\end{displaymath}
$\theta_i$ will be solutions of our version of the BA-equations.
 They will depend on the energy $E$, more specifically on
 which {\em block} {$\epsilon_i$} we are considering.

 Translational invariance now specifies:
\begin{equation}
\label{trans}
  a(x,y)=(e)^{x} a(\Delta),\;\;
  b_i(x,y)=(e)^{x} b_i(\Delta),\;\;
   c(x,y)=(e)^{x} c(\Delta),
\end{equation}
where $\Delta=y-x$, $e_n=\exp(\imath \theta_n)$ and $e=e_1e_2$.
Periodic boundary conditions require then:
\begin{displaymath}
a(\Delta)=(e)^\Delta a(N-\Delta),\;\;c(\Delta)=
(e)^\Delta c(N-\Delta),
\end{displaymath}
\begin{equation}
    b_2(\Delta)=(e)^\Delta b_1(N-\Delta),\;\;
    b_1(\Delta)=(e)^\Delta b_2(N-\Delta), \\
\label{pbc}
\end{equation}
meaning that the total momentum $P$ satisfies $P=\theta_1 +
 \theta_2 = 2l\pi/N$, where $l$ is integer.

Let us take the block $\epsilon_1=\epsilon_2=1$ first.
We try to build 2-pseudoparticle eigenstates out of
translationally invariant products of 1-pseudoparticle
 excitations at $x$ and $y$ with a weight function
 $D(x,y)$. Actually, as in the usual BA, we take advantage
 of the symmetry
 $E(\theta_1,\theta_2)=E(\theta_2,\theta_1)$ to write:
\begin{eqnarray}
|\theta_1,\theta_2>
   & =&\;\;\;\sum_{x<y} D_1(x,y) [ \gamma^*_1\, |x_{\,-}> +\;
|x_{\,0}>][\gamma^
   *_2\; |y_{\,-}> +\; |y_{\,0}>]\nonumber \\
   &   &+\sum_{x<y} D_2(x,y) [ \gamma^*_2\,|x_{\,-}> +\;
|x_{\,0}>][\gamma^*_1\,
    |y_{\,-}> +\; |y_{\,0}>].\\\nonumber
\label{ans2}
\end{eqnarray}
where $\gamma_i=\gamma(\theta_i),\gamma_i^*=\gamma(-\theta_i)$.
 Comparing this with equ.(\ref{2par}) and using
equs.(\ref{trans}), we have the following parametrisation:
\begin{displaymath}
a(\Delta)=\gamma_1^* \gamma_2^*\; D(\Delta),\;\;
c(\Delta)=D(\Delta),
\end{displaymath}
\begin{equation}
b_1(\Delta)=\gamma_1^*e_2^{\Delta} + \gamma_2^*\,e_2^N
 e_1^{\Delta} ,\;\;\;
b_2(\Delta)=\gamma_2^*e_2^{\Delta} + \gamma_1^*\,e_2^N
e_1^{\Delta} ,
\label{sp2}
\end{equation}
where $D(\Delta)=e_2^{\Delta}D_1(\Delta)+e_2^N
 e_1^{\Delta}D_2(\Delta)$ and $D_2(\Delta)=
e_2^N D_1(N-\Delta)$.
$H$ may now be applied to $|\theta_1,\theta_2>$ to
obtain a  set of coupled equations for $a(\Delta),b_i(\Delta) $ and
$c(\Delta)$.
If the pseudoparticles are not neighbors, i.e. between
any state $|->$ or $|0>|0>$ there is always a state $|+>$,
 then they do not interact and the state $|\theta_1,
\theta_2>$ will automatically be an eigenstate with
energy $E=2+ \gamma_1\gamma_1^*+\gamma_2\gamma_2^*$.
 The real problem arises of course, when pseudoparticles
are neighbors, so that they interact and we have no
guarantee, that the total energy is a sum of single
 pseudoparticle energies.

Therefore, following ref.\cite{P}, we split the
equations into {\em far} equations, when $\Delta>\Delta_{min}$
and {\em near } equations, when $\Delta=\Delta_{min}$.
 Here $\Delta_{min}=1$ for $a(\Delta),b_1(\Delta)$ and
$\Delta_{min}=2$ for $b_2(\Delta),c(\Delta)$.
The far equations are:
\begin{eqnarray}
\label{feq}
Ea(\Delta) &=& 2(\beta+1/\beta)a(\Delta)+(1+1/e)a(\Delta +1)
+(1+e) a(\Delta -1 )  \nonumber \\
       & &       + \sqrt\beta (b_1(\Delta-1) +
 b_2(\Delta +1)/e )
                  +1/\sqrt\beta (b_1(\Delta) +
b_2(\Delta)),\\ \nonumber\\
Eb_1(\Delta)&= &(1+\beta + 1/\beta)b_1(\Delta) +
 b_1(\Delta +1)/e +b_1(\Delta -1 ) e \nonumber \\
        &  & +\sqrt\beta a(\Delta +1)+1/\sqrt\beta a(\Delta)
 +\sqrt\beta/e c(\Delta +1)+1/\sqrt\beta c(\Delta),\\
    \nonumber\\
Eb_2(\Delta)&= & (1+\beta + 1/\beta)b_2(\Delta) +
 b_2(\Delta -1)+b_2(\Delta +1 )\nonumber\\
        &  &   + \sqrt\beta e a(\Delta-1)+
1/\sqrt\beta a(\Delta) +\sqrt\beta c(\Delta -1)+
1/\sqrt\beta c(\Delta),\\    \nonumber\\
Ec(\Delta)&=& 2c(\Delta)+1/\sqrt\beta (b_1(\Delta)
 + b_2(\Delta) ) +\sqrt\beta (e b_1(\Delta -1)
+b_2(\Delta +1))
\label{feqc}.
\end{eqnarray}

These we already know   to be satisfied by our ansatz.
 In fact a sufficient conditions for the far equations
 to be true is:
\begin{eqnarray}
D_i(\Delta)(\beta+1/\beta) + D_i(\Delta +\zeta_i)/e_i +
D_i(\Delta-\zeta_i)e_i&=
   &
     \gamma_i \gamma_i^*D_i(\Delta),\nonumber\\
\sqrt\beta D_i(\Delta +\zeta_i)/e_i +
1/\sqrt\beta D_i(\Delta)&=&\gamma_i^* D_i(\Delta),\\
\sqrt\beta D_i(\Delta-\zeta_1)e_i +
1/\sqrt\beta D_i(\Delta)&=&\gamma_i D_i(\Delta),\nonumber,
\end{eqnarray}
where $\zeta_1=1,\zeta_2=-1$.
If we use equs.(\ref{sp2}) and put $D_i(\Delta)=const_i$,
 they do indeed hold.

Now turn to the near equations:
\begin{eqnarray}
E a(1)&= &(\beta+1/\beta)a(1)+(1+1/e)a(2)+
1/\sqrt\beta b_1(1) +\sqrt\beta/e b_2(2),\\
\nonumber\\
E b_1(1)&=& (1+\beta)b_1(1) + b_1(2)/e+
\sqrt\beta a(2)+1/\sqrt\beta a(1)+\sqrt\beta c(2)/e,\\
\nonumber\\
\label{n}
E b_2(2) &=& (1+1/\beta)b_2(2)+b_2(3)+\sqrt\beta e a(1)
+1/\sqrt\beta a(2)+1/\sqrt\beta c(2),\\
\nonumber\\
E c(2) &=&3c(2) +\sqrt\beta
(eb_1(1)+b_2(3))+1/\sqrt\beta(b_1(2)+b_2(2))\nonumbe
   r\\
       & &+ c(2)(\beta+1/\beta)/(E-\beta-1/\beta)\label{nc}.
\end{eqnarray}

Now we notice two features implied by equs.(13),(15). First,
the wavefunctions  develop a 'discontinuity'. To see this,
take the far equation\footnote{
            The far equations are satisfied for any value of
 $\Delta$, but the function $c(\Delta)$ of equ.(13a) for
 example, equals the wavefunction only for
 $\Delta>\Delta_{min}$.} for $c(\Delta)$,
 put $\Delta=2$ and subtract from it the near equation
 equ.(\ref{nc}) for $c(2)$. We get:
\begin{equation}
c(2)(1 + \frac{\beta+1/\beta}{E-\beta-1/\beta})=0.
\label{ezero}
\end{equation}
This requires either $c(2)=0$ or $E=0$, both of them
unacceptable in the block $\epsilon_1=\epsilon_2=1$.
 Thus $c(2)$ in equ.(\ref{nc}) is not the continuation of
 $c(\Delta)$ from $\Delta>2$.( Except possibly for $E=0$,
 which is in fact the case.)
Second, the choice  $D_1(\Delta)=const$, which would be the
usual BA, does in fact NOT solve our near equations for any
 choice of $const$.

 Therefore we proceed as follows\cite{P}.
Let us  make a minimum modification, maintaining
 $D_1(\Delta)=const$ for $\Delta>\Delta_{min}$ and
leaving $a(1),b_1(1),b_2(2),c(2)$ as arbitrary constants.
 This means that we modify the ansatz equ.(\ref{sp2})
only when  pseudoparticles do
interact. Hence, we set:
\begin{equation}
\label{an3}
a(\Delta) =  \gamma_1^* \gamma_2^*\,Q(\Delta),\;\;
c(\Delta) =  Q(\Delta),\;\;\Delta>2
\end{equation}
\begin{equation}
b_1(\Delta)=\gamma_1^* e_2^{\Delta} + \gamma_2^* e_1^{\Delta}e_2^N,\;\Delta>1,
\label{ans3}
b_2(\Delta)= \gamma_2^* e_2^{\Delta}+
 \gamma_1^* e_1^{\Delta}e_2^N,\;\;
\end{equation}
\begin{displaymath}
 a(1)=a_1,\;\;
b_1(1)=b_1,\;\;
b_2(2)=b_2,\;\;
c(2)=c_2.
\end{displaymath}
Here  $Q(\Delta)=e_2^{\Delta}+e_1^{\Delta}e_2^N$ and
 $a_1,b_1,b_2$ and $c_2$ are constants to be determined.

All equations can now be satisfied \cite{KL} - {\em provided}
 the following consistency condition holds:
\begin{equation}
   e_2^N=-\frac{e_2}{e_1}\;\;
        \frac{(1+1/e)e_2-\epsilon}{(1+1/e)e_1-\epsilon},
\label{cons}
\end{equation}
where $ \epsilon = E-(q+1/q)$ and  we used
 $\beta+1/\beta+1=q+1/q$.
The unknown constants are given by:
\begin{equation}
a_1=\tilde a(1) + \frac{1}{e}F,\;\;\;
b_1=\tilde b_1(1) - \frac{1}{\sqrt\beta e}F,\;\;\;
b_2=\tilde b_2(2) - \sqrt\beta F,\;\;\;
c_2=\tilde c(2) + F.
\label{neart}
\end{equation}
Here $ \tilde a(1),\tilde b_1(1),\tilde b_2(2) ,\tilde c(2)$
 are the functions defined in equs.(\ref{an3}),(\ref{ans3})
 continued to $\Delta=\Delta_{min}$ and $ F=c_{\,0}Q(1)e/(1+e)$.

 In the block $\epsilon_1=\epsilon_2=1$, this can be rewritten
as:
\begin{equation}
   e_2^N=-\frac{1+e+(q+1/q)e_2}{1+e+(q+1/q)e_1}.
\label{cons1}
\end{equation}
This is the same consistency equation one finds for the
$6$-vertex model and  reduces to that of ref.\cite{P} in
 the undeformed case $\lambda=0$.
%The possibility of satisfying all equations is of course
a consequence of the %underlying Hecke algebra or
Yang-Baxter equations.

If we repeat the same procedure  for other blocks, we have
to change the $1$-particle wavefuncions. For $E\neq 0$,
equs.(\ref{neart}) remain identical and so does the
consistency condition equ.(\ref{cons}), provided we use
 the correct energy: $E = E(\epsilon_1,\epsilon_2)$.
 Therefore we get different BA equations for every block
$(\epsilon_1,\epsilon_2
   )$.

Only the case $E=0$ falls out of line, as we may expect
from equ.(\ref{ezero}). In this case the consistency condition
 is that of a free theory:
\begin{equation}
   e_1=e_2\,,\;\; e_i^N=1.
\end{equation}
There is a set of degenerate eigenfunctions, one of them
 continuous as expected.

%This same condition is met for $r=3/2$, where we scatter
% a moving pseudoparticle with a stationary {\em impurity}.

The generalization to any $r$ is straightforward. Since
the Yang-Baxter equations are satisfied, there is
only 2-pseudoparticle scattering ( - if we use
 S-matrix language~).
 Therefore near equations, where more then 2 pseudoparticles
 become neighbors, will not give any new
restrictions\cite{KL}\cite{SW}. The $n$-pseudoparticle phase
shift will be a sum of $2$-pseudoparticle phase shifts
$\Theta(\theta_1,\theta_2
   )$:
\begin{equation}
\Theta(\theta_1,\theta_2)=-\imath\log\left(
               -\frac{e_2}{e_1}\;\;
        \frac{(1+1/e)e_2-\epsilon}{(1+1/e)e_1-\epsilon}
                  \right),\; E\neq 0 .
\end{equation}

The ground state  of $(-H)$ lies in the sector $r=[N/2]$,
where $[z]$ stands for the greatest integer smaller or equal
to $z$. If $N$ is even, this sector contains the state
$\prod_x |0,x>$. For periodic boundary conditions $(-H)$
applied to this state creates an extra contribution to the
energy arising from the $S_z=0$ - states  at position $x=N$
and $x=0$. To avoid this complication, we restrict $N$ to be
odd. The BA equations for $E\neq 0$ are then given by:
\begin{equation}
     e_a^N=\prod_{b=1,b\neq a}^r\;
               -\frac{e_a}{e_b}\;\;
\frac{(1+1/(e_ae_b))-\epsilon}{(1+1/(e_ae_b))e_b-\epsilon}
\;\;
                            a=1,\ldots,r.
\label{baeq}
\end{equation}

The above results may be extended to other boundary conditions,
 like twisted and free ones\cite{KL}. The free boundary
condition BA equations are:
\begin{equation}
\label{free}
e_a^N=\prod_{b=1,b\neq a}^r\;\frac{B(1/e_a,e_b)}
{B(e_a,e_b)},\;\;
              a=1,\ldots,r
\end{equation}
where
\begin{equation}
B(e_a,e_b)=\frac{e_b}{e_a}[e_b + 1/e_a - \epsilon]
                          [1/e_b+1/e_a - \epsilon].
\end{equation}
 For twisted boundary conditions and $N$ = odd, the
$\epsilon_i=1$ BA equations differ from the $6$-vertex
 equations only by the replacement of
$\Phi\rightarrow 2\Phi$.
%
%conclusion

Our spin $1$ hamiltonian is hermitian only for $q+1/q>2$
and outside this region complex eigenvalues arise.
Nevertheless the block  $\epsilon_i=1$, which contains
the ground state of the complete model, constitutes by
 itself a perfectly unitary theory with real eigenvalues.
 This shows, that a theory which is not unitary, may
 nevertheless possess a subset of eigenvalues, belonging
to a unitary one, although the corresponding states are
not decoupled from the rest. A more detailed investigation
 of this point should also shed light on similar situations
in other settings\cite{SM}.

Our results are in agreement with the rather extensive
numerical calculations performed by Alcaraz and
 Malvezzi\cite{AM}, who in particular observed that
 $(-H)$  and the $6$-vertex hamiltonian share the same
 ground-state for $N$ = odd. Thus  e.g. both have conformal
 anomaly $c=1$.

We may also verify, that our BA equations for free
boundary conditions equ.(\ref{free}), are highest
weight states of the quantum group $U_qsl(2)$. For
details we refer to our forthcoming paper\cite{KL}.
%%%%

Let us summarize our results.
 In order to implement an exact solution, the hamiltonian
density had to satisfy the Hecke algebra and due to
$U(1)$-invariance, there existed a reference state
$|\Omega>$, satifying $H|\Omega>=0$.
Pseudoparticles are now created on top $|\Omega>$ and the
 machinery of the coordinate BA is applied.
This requires the introduction of 'discontinuous'  wave
functions, which permits us to obtain the eigenstates
by a generalizations of the usual coordinate BA.
 Even
when our hamiltonian is not hermitian, it always has a
block of real eigenvalues, identical to the ones of the
6-vertex model.
Finally, questions like:  do the Yang-Baxter equations
guarantee, that our procedure always works; do the BA
 solutions constitute a complete set of states; what,
 if any, is the consequence of the 'discontinuity' in
a model on a discrete chain etc.
 is left for a future publication, where we will also
treat other models by the same method\cite{KL}.

We have the pleasure to thank Dr. A. Malvezzi for
 informative discussions.

%******************************************************************************
%


\begin{thebibliography}{99}
%
\bibitem{YB} Yang C.N., {\em Phys. Rev. Lett.} {\bf19}, 1312, (1967).
              Baxter R.J.,{\em Exactly Solved Models
in Statistical Mechanics}
              ( Academic Press, London, 1982 ).
\bibitem{B} Bethe H. A., {\em Z. Physik } {\bf 71}, 205, (1931).
\bibitem{AKL} Alcaraz F.C., K\"oberle R. and Lima - Santos A.,
 {\em Int. J. Mod. Phys.} {\bf A7}, 7615, (1992).
\bibitem{TF} Takhtadjian and L. Fadeev,
               {\em Russ. Math. Surveys } {\bf 34}, 11, (1979).
\bibitem{P} Parkinson J.B., {\em J.Phys. C.: Solid State
             Phys.} {\bf 21}, 3793, (1988).
\bibitem{BB} Barber M. and Batchelor M.F., {\em Phys. Rev.} {\bf B40}, 4621,
(19
   89).
\bibitem{TL} Temperley H. N. V. and Lieb E. H., {em Proc.
R. Soc. London Ser.} {\em A322}, 251, (1971).
\bibitem{K} Kl\"umper A., {\em Int. J. Mod. Phys.} {\bf B4}, 171, (1990).
\bibitem{BMNR} Batchelor M.T.,Mezincescu L.,Nepomechie R.I.,
 Rittenberg V., {\em J. Phys.} {\bf A23}, L141, (1990).
\bibitem{KR} Kirillov and Reshitkin {\em J. Phys} {\bf A },
  ,(199 ).
\bibitem{KL} K\"oberle R. and Lima - Santos A.,
to be published.
\bibitem{SW} Shankar R. and Witten E., {\em Phys. Rev.} {\bf D17}, 2134,
(1978).
    %\bibitem{RB} Baxter R. J.{\em J. Stat. Phys.} {\bf 28}, 1, (1982).
\bibitem{SM} Smirnov F. A.{\em Int. J. Mod. Phys. } {\bf A6}, 40, (1991).
\bibitem{AM}  Alcaraz F. C. and Malvezzi A.{\em J. Phys.} {\bf A25}, 4535,
(1992
   ).
\end{thebibliography}
\end{document}